\documentclass[prd,preprintnumbers,amsmath,amssymb,nofootinbib,a4paper]{revtex4}
\usepackage{epsfig}
\usepackage{multirow}

\begin{document}


\title{Reply to ``Commentary on ``Total Hadronic Cross Section Data and the 
Froissart-Martin Bound", by Fagundes, Menon and Silva"}

\author{D.A. Fagundes, M.J. Menon, P.V.R.G. Silva}

\affiliation{Universidade Estadual de Campinas - UNICAMP\\
Instituto de F\'{\i}sica Gleb Wataghin \\
13083-859 Campinas, SP, Brazil \\
fagundes@ifi.unicamp.br, menon@ifi.unicamp.br, precchia@ifi.unicamp.br}

\begin{abstract}
A  reply to the above mentioned commentary by M.M. Block and F. Halzen
on our quoted work is presented. We answer to each point raised by these authors
and argument that our data reductions, strategies and methodology
are adequate to the nonlinear-fit-problem in focus. In order to exemplify 
some arguments, additional information 
from our subsequent analysis is referred to. A brief commentary on the
recent results by Block and Halzen is also presented.
We understand that
this reply gives support to the results and conclusions presented in our quoted work. 
\end{abstract}

\maketitle


\vspace{0.7cm}

\textbf{Table of Contents}

\vspace{0.2cm}

I. Introduction

\vspace{0.2cm}

II. Reply to the Criticisms

\vspace{0.1cm}

\ \ \ II.A On the $\rho$ Information

\vspace{0.1cm}

\ \ \ II.B Statistical Probabilities

\vspace{0.1cm}

\ \ \ II.C Prediction of the 7 TeV Total $pp$ Cross Section

\vspace{0.2cm}

III. Discussion

\vspace{0.1cm}

\ \ \ III.A On the FMS Analysis and Results

\vspace{0.1cm}

\ \ \ III.B Brief Commentary on BH Analysis and Results

\vspace{0.2cm}

IV. Conclusions and Final Remarks

\vspace{0.7cm}

\section{Introduction}
\label{s1}


In a recent communication M.M. Block and F. Halzen \cite{bhcomm} 
(hereafter referred to as BH) have presented some
critical comments  on our analysis \cite{fms1} (hereafter FMS). Some points raised 
by these authors have already been
addressed and discussed in our subsequent work \cite{fms2} (hereafter FMS2),
available as arXiv since August 16, 2012.
In this note we reply to the BH criticisms directed to FMS,
recalling also some aspects of interest presented in FMS2.

First, to facilitate the discussion and the reference to each part of
the BH commentaries, some explanations on the organization of this reply are in order.
Excluding the acknowledgments and references,
BH arXiv comprises four pages and the effective criticisms to FMS correspond to approximately
one page. All the rest of the manuscript (three pages) largely overlap with their
previous work \cite{bh12} (as advised in the ``arXiv admin note" \cite{bhcomm}).
We shall focus on this 25 \% material, in our section 
\ref{s2}.
Although not characterized as criticisms, the rest of the BH reproduces
their former work on the subject, as a kind of lesson to be learned.
In this respect, a discussion on some aspects of the FMS analysis and a brief commentary 
on the BH former work
are presented
in section \ref{s3}. Our conclusions and final remarks are the contents of section \ref{s4}.

\section{Reply to the Criticisms}
\label{s2}

The content of the criticisms to FMS, presented in BH, can be divided
in three blocks, one block referring to the $\rho$ information (page 1 in BH), another block
referring to statistical probabilities (page 2) and the last one to predictions at 7 TeV
(page 2). In what follows, each block will be treated as a subsection, in which we first situate
and summarize the commentary, or quote it explicitly, and then present our reply.

\vspace{0.2cm}

\subsection{On the $\rho$ Information}

\vspace{0.2cm}

\textit{- Commentary}

\vspace{0.2cm}

The first effective criticism appears in page 1, fourth paragraph of the section \textit{Introduction}.
It concerns the fact that in FMS the $\rho$ information was not used in
global fits with the total cross section data. According to them, ``a statement alluded
to (but \textit{not} carried out) in Appendix...". They also add:
``... in their Appendix, they give a rather cumbersome evaluation using their Variant 3 model, 
to \textit{separately evaluate} $\rho$...."

\vspace{0.3cm}

\textit{- Our reply}

\vspace{0.2cm}

In FMS, the analysis has been based only on the $\sigma_{tot}$ data
(without the inclusion of the $\rho$ information) for the six reasons explained
there, which we consider as six facts.
However, addressing the comments by one of the \textit{three referees}
in the submission to the Braz. J. Phys., we have included Appendix A in a revised version. 
In this appendix
we have shown that, even in the case of the largest values of the
exponent $\gamma$ (Method 1, V3 and Method 2, V5), the \textit{predictions} for  $\rho(s)$ are 
in agreement
with the experimental information. 

To connect $\sigma_{tot}(s)$ and $\rho(s)$ in an analytical way, we have used singly-subtracted
derivative dispersion relations in the operational form introduced by
Kang and Nicolescu \cite{kn} (also discussed in \cite{am}). In particular we have obtained an
\textit{extended} analytical result for the case of $\gamma$ as a \textit{real} parameter (equations A7 and A8
in Appendix A of FMS). 

In respect to the effect of the $\rho$
information in data reductions, we have stated 
at the end of Appendix A
(the references that follows concern FMS): ``Finally, we recall that in simultaneous fit
to $\sigma_{tot}$ and $\rho$  the subtraction constant affects both the low- and 
high-energy regions
[47,48].
That is a consequence of the strong correlation among the subtraction constant and all
the other physical free fit parameters. We plan to discuss this consequence
and other aspects of the fit procedures in a
forthcoming paper." Also, in the last paragraph of our
conclusions
(third sentence) we added: ``These are aspects
that we expect to consider in a future work, since they may provide information
that is complementary to the results here presented."

In fact, in the FMS2  we have extended the FMS
analysis in several aspects as referred to and outlined in the introduction of FMS2.
In special, not only individual but also novel simultaneous fits to total cross section
and $\rho$ information have been developed, leading to solutions with $\gamma$ 
greater than 2, \textit{despite the constraint involved}. The origin and role of the subtraction constant
have also been discussed in some detail.

Therefore, we see no sense in depreciating statements  like ``alluded but not carried out"
or ``they give a rather cumbersome evaluation".

\vspace{0.2cm}

\subsection{Statistical Probabilities}

\vspace{0.2cm}

The second criticism in BH appears in page 2, section \textit{Statistical Probabilities}
and involves five paragraphs, four with criticisms (left column) and the final one with 
the conclusions (right column).
The main point here concerns the use of the integrated probability $P(\chi_0^2;\nu)$
to punctually analyze the reliability of the fit results. We shall treat each paragraph 
separately and in sequence. However, before that, some aspects on our fit
procedures and on statistical analysis demand a few comments for further reference.

\vspace{0.3cm}

$\bullet$ Introductory remarks

\vspace{0.2cm}

- \textit{On the fit procedures}

\vspace{0.1cm}

The parametrization for the total cross section used in FMS, introduced by Amaldi \textit{et al.} \cite{amaldi}, reads
\begin{eqnarray}
\sigma_{tot}(s) =  
a_1\, \left[\frac{s}{s_l}\right]^{-b_1} + 
\, \tau \, a_2\, \left[\frac{s}{s_l}\right]^{-b_2} +
\alpha\, + \beta\, \ln^{\gamma} \left(\frac{s}{s_h}\right),
\end{eqnarray}
\label{eq1}

\noindent
where $\tau$ = -1 (+1) for $pp$ ($\bar{p}p$) scattering and $s_l$ = 1 GeV$^2$ is fixed.
The dependence
is linear in four parameters ($a_1$, $a_2$, $\alpha$ and $\beta$) and nonlinear
in the other four ($b_1$, $b_2$, $\gamma$ and $s_h$). As stated by 
Bevington and Robinson \cite{bev} (section 8.2 Searching Parameter Space) and also quoted recently in \cite{fmg},

\begin{quote}
``Fitting nonlinear functions to data samples sometimes seems to be more an
art than a science."
\end{quote}

Nonlinear data reductions are not a trivial task. They demand
a methodology for the choice of the initial (feedback) values of the free parameters.
Our strategy has been to test a \textit{grid} of different
(physical) feedback values for the free fit parameters so as to check the stability of the results,
as shortly recalled in what follows.

In both cases, FMS and FMS2, we have considered as feedback the
results already found by the Particle Data Group (PDG), which uses the standard COMPETE 
Collaboration highest-rank parametrization
($\gamma$ = 2, a fixed parameter) \cite{compete1,compete2}. In FMS we used the 2010 PDG edition \cite{pdg10} and in FMS2 the updated result 
from the 2012 PDG edition \cite{pdg12}. 
Although applied to only a \textit{subset} of the dataset analyzed in PDG,
we understand that with these ``conservative choices" ($\gamma$ = 2), we start with 
reasonable stable
solutions (already found by the PDG). With this input we are then able to investigate possible \textit{departures from this solution} in the case of
$\gamma$ as a free parameter (including or not the TOTEM datum \cite{totem}, as done in FMS and including this
point in all fits, as done in FMS2). \textit{In addition}, in order to investigate the effect of the feedback values in
the fit results, we have considered another distinct choices in both FMS and FMS2 and 3 different versions
in each case of FMS (referred to as six variants and a direct fit).

As explicitly quoted in FMS and FMS2, our data reductions have been carried out with the objects of the 
class TMinuit of the ROOT Framework \cite{root}. 
The statistical interpretations of the fit results, as well as aspects related to the error matrix, correlation matrix 
and analytical error propagation, have been based on the Bevington-Robinson book \cite{bev}.

For further reference, we recall that the error (or covariance) matrix
provides the variances (diagonal) and covariances (out of diagonal)
associated with each free fit parameter. The symmetric correlation matrix gives
a measure of the correlation between each pair of free parameter
through a coefficient with numerical limits $\pm$ 1 (full correlation)
and 0 (no correlation).

In the minimization program \cite{minuit1,minuit2,minuit3} a Confidence Level of one standard deviation
was adopted in all fits (UP = 1). In each test of fit, successive running of the Migrad
have been considered (up to 200 calls in FMS and up to 2,000 calls in FMS2), 
until full convergence has been reached, with the smallest FCN ($\chi^2$)
and EDM (Estimated Distance to Minimum), specifically EDM $< 10^{-4}$ (adequate for the 
one sigma CL). In addition, the error in the parameters should not exceed the central value.

Among \textit{several different tests}, we have \textit{selected}, under the above criteria, 
the seven variants presented
in FMS. These variants are related to two different choices for the input values
for all fits, denoted Method 1 and Method 2 in FMS.
For further reference and clarity, we summarize and situate below the structure of the
grid considered in FMS (V stands for variant and DF for direct fit, also a variant).

\vspace{0.2cm}

\noindent
Method 1.
Initial feedback values from PDG 2010 (Table 1, second column):
\begin{equation}\nonumber
\left.\begin{array}{rl}
\ \ \sqrt{s}_{max} \text{\ =\ 1.8\ TeV\ Ensemble\ (Table\ 1)} \\
                                                  \\
\sqrt{s}_{max} \text{\ =\ 7\ TeV\ Ensemble\ (Table\ 2)\ }
\end{array}\right\}
\text{\ variants\ DF, V1, V2, V3}
\end{equation}

\vspace{0.2cm}

\noindent
Method 2.
Initial feedback values distinct from Method 1 (Table 3, second column):
\begin{equation}\nonumber
\left.\begin{array}{rl}
\ \ \sqrt{s}_{max}\text{\ =\ 1.8\ TeV\ Ensemble\ (Table\ 3)} \\
                                                  \\
\sqrt{s}_{max}\text{\ =\ 7\ TeV\ Ensemble\ (Table\ 4)\ }
\end{array}\right\}
\text{\ variants\ V4, V5, V6}.
\end{equation}

\vspace{0.2cm}

In the six variants (including the DF) we have investigated the effects of fixing or not
the three fundamental parameters directly related with the energy dependence, namely
$b_1$, $b_2$ and $\gamma$. For the Reggeon intercepts we have tested either
\textit{ ad hoc} fixed values 1/2 or fixed (central) values from spectroscopic
data (Chew-Frautschi plots).

We stress that these results constitute final solutions, selected
under the above mentioned criteria.
Therefore our strategy in FMS (and FMS2) did not involve extremely detailed use of
different routines to possibly reach an absolute minimum.
Our point has been to investigate a \textit{grid} of reasonable
physical choices for feedbacks and variants for two ensembles, obtaining solutions 
through standard running of the MINUIT.
Despite of these possible limitations, we have found several reasonable consistent solutions
with $\gamma$ greater than 2 and that has been the only essential point raised in FMS (and 
developed also in FMS2).

\vspace{0.2cm}

- \textit{On statistics}

\vspace{0.1cm}

In FMS and FMS2, following the PDG procedure, 
the dataset include statistical and \textit{systematic errors}, added in quadrature.
In our opinion, the inclusion of systematic errors puts certain
limits in a \textit{full} statistical interpretation of the fit results.

In fact,
the $\chi^2$ test for goodness of fit is based on the assumption
of a \textit{Gaussian error distribution} \cite{bev}. Although statistical 
uncertainties are considered to follow this distribution, 
that, certainly, is not the case for systematic uncertainties, which are equally probable
quantities. Therefore, we understand that a \textit{full} statistical interpretation of data reductions
including systematic uncertainties
has a somewhat limited validity, specially in what concerns integrated probability (due to the
inclusion of equally probable quantities).

In the FMS and FMS2,
the corresponding DOF ($\nu$) and $\chi^2/\nu$ for each fit has been displayed only
to shown that they constitute reasonable (acceptable) statistical results. The 
condition of reduced $\chi^2$ closest to 1.0
has been one of the criteria used to select a given result, but not the only one.
Attempts to a full statistical interpretation of the results, mainly in terms
of integrated probabilities,
may lead to questionable conclusions, as discussed in what follows.

At last, it is important to note that the focus in FMS does not concern comparison
among models (or variants),
but between two ensembles, with or without the TOTEM datum. Also, when we refer to \textit{statistically 
consistent results} for $\gamma$ = 2 or $\gamma$ above 2, we mean that \textit{the corresponding numerical
result for $\gamma$ is consistent within their uncertainty}.

Let us now treat each paragraph from section \textit{Statistical Probabilities} in BH.
We shall adopt here their notation (lower index 1 for the $\sqrt{s}_{max}$ = 1.8 TeV
ensemble and lower index 2 for the $\sqrt{s}_{max}$ = 7 TeV ensemble).

\vspace{0.3cm}

$\bullet$ First Paragraph

\vspace{0.3cm}

- Commentary

\vspace{0.2cm}

BH discuss our results in Table 1 of FMS ($\sqrt{s}_{max}$ = 1.8 TeV),
comparing the data reductions that can be summarized as follows:

\vspace{0.1cm}

DF$_1$: $\quad$ $\nu$ = 156, $\quad$ $\chi^2/\nu$ = 0.931, $\quad$ $P_{DF_1}$ = 0.721,$\quad$ case of $\gamma$ = 2 (fixed) 

\vspace{0.1cm}

$\ \,$V1: $\quad$  $\nu$ = 155, $\quad$ $\chi^2/\nu$ = 0.937, $\quad$ $P_{V1}$ = 0.701, $\quad$ case of $\gamma$ free 

\vspace{0.1cm}

 According to Block and Halzen:
``... we get the somewhat strange result that FMS have a \textit{...better, somewhat
more reliable} fit when they fix the value of $\gamma$ at 2,..., than they allow it to float,
suggesting perhaps that the \textit{true} minimum $\chi^2$ was not achieved in their minimization process."
They state in the last sentence: ``In any event, FMS concluded that the value $\gamma$ = 2 was correct for the energy interval 5 $\leq \sqrt{s} \leq$ 1,800 GeV."

\vspace{0.3cm}

- Our reply:

\vspace{0.2cm}

The V1 result has been obtained from DF$_1$ as feedback.
Based on Bevington and Robinson \cite{bev} (Chapter 11) and in our above comments on statistics, 
for $\nu \sim$ 155/156 we do not
think that
$P_{DF_1} \sim$ 0.72 $>$  $P_{V1} \sim$ 0.70 strictly implies in a better fit,
namely that DF$_1$ could be ``more reliable" than V1. In fact, let us compare
the values of the free parameters in both fits in Table 1 of FMS (third and fourth
columns). Note that the central values of the parameters
are identical up to 3 figures, except for $\beta$ (0.264 and 0.263) and $s_h$
(12.0 and 12.2). All the parameters in both fits are consistent within their errors,
leading to the conclusion that both results are, effectively, equivalent
(the corresponding curves in Figures 1 and 2 of FMS overlap).

We understand that in this nonlinear fit,
based on the same dataset ($\sqrt{s}_{max}$ = 1.8 TeV), to let
free \textit{only one parameter} ($\gamma$) does not allow the punctual statistical interpretation by
Block and Halzen (we shall return to this  point in the discussion
on the fourth paragraph).
The very small differences in the central values of the parameters, and those in the errors,
are associated with the \textit{correlations among all the fit parameters},
resulting, in this particular case, in a small decrease of the probability when $\gamma$ is let free. 
The correlation coefficients associated with both fits are displayed in Table \ref{t1}:
DF$_1$ above the diagonal of the table and V1 below the diagonal (we shall return to this point in what follows).
Perhaps there can be some small differences in reaching a true minimum, but,
in our opinion, that does
not invalidate our results and interpretation.

\begin{table}[ht]
 \centering
\caption{Correlation coefficients from the (symmetric) correlation matrices \cite{bev,root} associated with 
DF$_1$ and V1$_1$ results. The off-diagonal coefficients from DF$_1$ are displayed above the diagonal
of the table (not filled) and those from V$_1$, below that diagonal.}
\begin{tabular}{|c c | c c c c c c c c|}
\hline
 &          & \multicolumn{8}{c|}{DF$_1$} \\
 &          & $a_1$  &  $b_1$ &  $a_2$ &  $b_2$ &  $s_h$ & $\alpha$ & $\beta$ & $\gamma$ \\
\hline
\multirow{8}{*}{\ V1$_1$\ }
 &   $a_1$  &        &  0.504 &  0.259 &  0.238 &  0.089 &   0.221  &  -0.090 &  - \\
 &   $b_1$  &  0.754 &        &  0.079 &  0.073 &  0.897 &   0.953  &   0.772 &  - \\
 &   $a_2$  &  0.187 &  0.034 &        &  0.989 &  0.024 &   0.021  &   0.024 &  - \\
 &   $b_2$  &  0.163 &  0.031 &  0.981 &        &  0.024 &   0.020  &   0.026 &  - \\
 &   $s_h$  & -0.363 & -0.059 &  0.072 &  0.071 &        &   0.987  &   0.966 &  - \\
 & $\alpha$ & -0.181 &  0.462 & -0.059 & -0.056 &  0.629 &          &   0.914 &  - \\
 & $\beta$  &  0.028 & -0.138 &  0.017 &  0.013 &  0.538 &  -0.116  &         &  - \\
 & $\gamma$ & -0.174 &  0.020 &  0.029 &  0.031 & -0.234 &   0.237  &  -0.912 &        \\
\hline
\end{tabular}
\label{t1}
\end{table}

Concerning the last sentence in the paragraph (quoted above), it is obvious
that we were not led to that conclusion 
based only in this particular case (DF$_1$ and V1), but in all the methods and variants
displayed in Tables I and III, as explained along the text in FMS (and summarized in the beginning
of section IV).

This first example already indicates the limits of a full statistical interpretation
in terms of the integrated probability. This interpretation, however, permeates all the paragraphs 
in BH, to be discussed next.

\vspace{0.5cm}

$\bullet$ Second Paragraph

\vspace{0.3cm}

Here they compare DF ($\gamma$ fixed) and V1 ($\gamma$ free) from Table 1 of FMS
($\sqrt{s}_{max}$ = 1.8 TeV) and Table 2 ($\sqrt{s}_{max}$ = 7 TeV).
Let us treat the two cases (DF and V1) separately.

\vspace{0.2cm}

- Commentary

\vspace{0.2cm}

In the case of DF ($\gamma$ fixed), summarizing the results,

\vspace{0.1cm}

DF$_1$: $\quad$ $\nu$ = 156, $\quad$ $\chi^2/\nu$ = 0.931, $\quad$ $P_{DF_1}$ = 0.721,$\quad$  $\sqrt{s}_{max}$ = 1.8 TeV

\vspace{0.1cm}

DF$_2$: $\quad$  $\nu$ = 157, $\quad$ $\chi^2/\nu$ = 0.930, $\quad$ $P_{DF_2}$ = 0.725, $\quad$ $\sqrt{s}_{max}$ = 7 TeV

\vspace{0.1cm}

They conclude:

``Thus, if $\gamma$ = 2 is satisfactory for the low energy data,
it appears to be exactly the same level of confidence when we include
the Totem point."

\vspace{0.2cm}

- Our reply:

\vspace{0.3cm}

First, it is important to stress that DF$_2$ has been obtained with DF$_1$ as feedback.
Since $\gamma$ = 2 is fixed and only the TOTEM point has been added, it is not
expected a drastic change in the fit results in terms of $\chi^2/\nu$
(and possibly in terms of integrated probability) as is the case.
However, note that the \textit{high energy parameters} ($\alpha$, $\beta$
and $s_h$) are not identical up to 2 figures. In particular, the value of $s_h$ in
DF$_1$ is two times the value in DF$_2$.
As a consequence, from Fig. 1 of FMS, the high-precision TOTEM result is not adequately described in both cases: the curve from DF$_1$ ($\sqrt{s}_{max}$ = 1.8 TeV) lies below the lower
error bar and even DF$_2$ ($\sqrt{s}_{max}$ = 7 TeV) lies only through the lower error bar.
At this point we agree that a solid conclusion demands the evaluation of the uncertainties in the
curves from error propagation (as we have done in FMS2). However, we do not think
this lack here invalidates our arguments (we shall return to this point in the
next section).
Therefore, we can not agree with the above quoted conclusion by Block and Halzen.

\vspace{0.2cm}

- Commentary

\vspace{0.2cm}

In the case of V1 ($\gamma$ free), summarizing the results,

\vspace{0.1cm}

V1$_1$: $\quad$ $\nu$ = 155, $\quad$ $\chi^2/\nu$ = 0.937, $\quad$ $P_{V1_1}$ = 0.701,$\quad$  $\sqrt{s}_{max}$ = 1.8 TeV

\vspace{0.1cm}

V1$_2$: $\quad$  $\nu$ = 156, $\quad$ $\chi^2/\nu$ = 0.935, $\quad$ $P_{V1_2}$ = 0.709, $\quad$ $\sqrt{s}_{max}$ = 7 TeV

\vspace{0.1cm}

Here, however, BH compare V1$_2$ with DF$_2$, concluding that, once
$P_{V1_2}$ = 0.709 $<$  $P_{DF_2}$ = 0.725, ``this result is somewhat strange,
since the logarithmic power $\gamma$ was let adjustable in the V1$_2$."

\vspace{0.2cm}

- Our reply:

\vspace{0.3cm}

The comparison is not adequate. As explained in the text and summarized in the legend
of Tables 1 and 2, each result in Table II has been obtained using as feedback the
corresponding results in Table 1. Specifically, DF$_2$ (Table 2) uses DF$_1$ (Table
1) as initial values in the data reductions and the same is true for V1$_2$
(Table 2) and V1$_1$ (Table 1) and so on.

Therefore, V1$_2$ and DF$_2$ have been obtained from different feedback values, which may imply in different
regions of the $\chi^2$ minimum (and that is the essence of our strategy,
as already discussed in the introductory remarks to this section). 

Here, it may be interesting to note that, if we accept the BH confidence on $P(\chi_0^2;\nu)$
and focus on V1 then

\vspace{0.1cm}

\centerline{$P_{V1_2} \sim$  0.71 $>$ $P_{V1_1} \sim$  0.70}

\vspace{0.1cm}

\noindent
would imply in a solution with $\gamma_{V1_2} \sim$ 2.10 $\pm$ 0.03
($\sqrt{s}_{max}$ = 7 TeV) more reliable than a solution with 
$\gamma_{V1_1} \sim$ 2.00 $\pm$ 0.03
($\sqrt{s}_{max}$ = 1.8 TeV), corroborating therefore our conclusion. That all seems a vicious circle.

Anyway, from these arguments we can not agree with the above quoted conclusion by Block and Halzen:
we see nothing ``somewhat strange" in our results.

\vspace{0.3cm}

$\bullet$ Third Paragraph

\vspace{0.3cm}

- Commentary

Here they insist in the punctual statistical interpretation, quoting now
two fit results, V4 and V5, with Method 2 (different initial values from Method 1).

\vspace{0.2cm}

- Our reply:

In this case, they correctly compare two results associated with the same feedback values
(first column in Table 3). However and once more, in our opinion,

\vspace{0.1cm}

\centerline{$P_{V4_1} \sim$  0.62 $>$ $P_{V5_1} \sim$  0.60}

\vspace{0.1cm}

\noindent
for $\gamma$ = 2 and $\gamma$ free, respectively, does not imply in a significant
difference. There is only \textit{one additional free parameter}, the same dataset and the resulting
\vspace{0.1cm}

\centerline{$\chi^2_{V4_1}$ = 152.154 and  $\chi^2_{V5_1}$ = 152.133}

\vspace{0.1cm}
\noindent
reinforces our opinion. We attribute the differences in the values of the
parameters to the correlations among them (5 free parameters in V4 and 6 free parameters in V5).
Also, and once more, perhaps there can be some small differences in reaching a true minimum, but that does
not invalidate our results and interpretation.

\vspace{0.3cm}

$\bullet$ Fourth Paragraph

\vspace{0.3cm}

- Commentary

``To illustrate this anomaly, we recall to the reader that
the difference between the V4$_1$ and the Direct Fit (DF$_1$)
models was that in the V4 model, the Regge powers $b_1$
and $b_2$ were fixed at 1/2, whereas they were allowed to
vary in the Direct Fit model, \textit{raising} the probability from
$P_{V4}$ = 0.616 to $P_{V6}$ = 0.721, as expected - the \textit{exact
opposite} of the V4 to V5 effect, where the logarithmic
power $\gamma$ was varied from 2. Clearly, we question their
minimization program, or their use of it."

\vspace{0.2cm}

- Our reply:

Here, there seems to be some errors (misprints?) along the text, leading to a complete misleading
paragraph. That was the reason why we have quoted here the complete paragraph.

If we understood, they are attempting to compare, once
more, different methods, namely DF (Method 1 - initial values from the COMPETE - PDG 2010,
given in Table 1) with V4 (Method 2 - different initial values, given in Table 3).
They refer to $P_{V4_1}$ = 0.616 and $P_{V6_1}$ = 0.721, but in the last case
it should be $P_{DF_1}$, which reads 0.721 and not $P_{V6}$ that reads 0.701.
The paragraph seems to us rather anomalous.

It just seems strange these possible errors concerning V4 and V6
in Table 3 because these variants constitute  very illustrative results.
They are interesting examples because V6 has been obtained with V4 as feedback.
The V4 has 3 fixed parameters, $b_1$, $b_2$ (low energies) and $\gamma$ (high energies)
and in V6 these \textit{three} parameters are let free.
They can be summarized as before:

\vspace{0.1cm}

V4: $\quad$ $\nu$ = 158, $\quad$ $\chi^2/\nu$ = 0.963, $\quad$ \textbf{$P_{V4}$ = 0.616}
$\quad$ ($b_1$, $b_2$, $\gamma$ fixed)

\vspace{0.1cm}

V6: $\quad$  $\nu$ = 155, $\quad$ $\chi^2/\nu$ = 0.937, $\quad$ \textbf{$P_{V6}$ = 0.701}
$\quad$ ($b_1$, $b_2$, $\gamma$ free)

\vspace{0.1cm}

Here, in this nonlinear fit, we have \textit{three fit parameters floating} 
resulting in an effective rising of the integrated probability.
The corresponding correlation coefficients for both fits are given
in Table \ref{t2}: V4 above the diagonal of the Table and V6 below that diagonal.

\begin{table}[ht]
 \centering
\caption{Same as Table I for the V4 result (above the diagonal) the the V6 result
(below the diagonal).}
\begin{tabular}{|c c | c c c c c c c c |}
\hline
 &          & \multicolumn{8}{c|}{V4$_1$} \\
 &          & $a_1$  &  $b_1$ &  $a_2$ &  $b_2$ &  $s_h$ & $\alpha$ & $\beta$ & $\gamma$ \\
\hline
\multirow{8}{*}{\ V6$_1$\ }
 &   $a_1$  &        &  -     &  0.156 &  -     & -0.965 &  -0.991  &  -0.889 & - \\
 &   $b_1$  &  0.744 &        &  -     &  -     &  -     &   -      &   -     & - \\
 &   $a_2$  &  0.181 &  0.034 &        &  -     & -0.087 &  -0.101  &  -0.065 & - \\
 &   $b_2$  &  0.158 &  0.031 &  0.981 &        &  -     &    -     &   -     & - \\
 &   $s_h$  & -0.381 & -0.051 &  0.074 &  0.072 &        &   0.988  &   0.972 & - \\
 & $\alpha$ & -0.193 &  0.471 & -0.058 & -0.055 &  0.632 &          &   0.929 & - \\
 & $\beta$  &  0.003 & -0.157 &  0.023 &  0.018 &  0.565 &  -0.102  &         & - \\
 & $\gamma$ & -0.153 &  0.050 &  0.022 &  0.024 & -0.282 &   0.228  &  -0.920 &   \\
\hline
\end{tabular}
\label{t2}
\end{table}

\vspace{0.3cm}

$\bullet$ Fifth Paragraph

\vspace{0.3cm}

- Commentary

Block and Halzen conclude there is no statistical evidence for $\gamma > $ 2.

\vspace{0.2cm}

- Our reply:

Based on all the results presented in Tables 1 to 4 of FMS,
we conclude that, in what concerns fits with the used parametrization and once the TOTEM 
data is included in the dataset,
we obtain statistical solutions consistent with $\gamma$ greater than 2.

\vspace{0.5cm}

$\bullet$ Final Comment

\vspace{0.3cm}

Concerning the noticed ``anomalous" decrease of the integrated probability
when one parameter ($\gamma$) is let free, we add the comment that follows.
The parameters $a_1$, $b_1$, $a_2$ and $b_2$ are related to the low 
and intermediate energy region
(5 GeV $ \leq \sqrt{s} \lesssim $ 100 GeV), where the number of data points is much larger then
that at higher energies (where the parameters $\alpha$, $\beta$, $\gamma$ and $s_h$ play
the central role). As a consequence of the correlation among all parameters, when $\gamma$ is let
free, the low-energy parameters may also be affected, leading, in some cases, to a less reliable fit result
due to the large number of data points at the lower energies. Therefore,
we understand that this possible ``anomalous" effect may not be connected with
the position of a true minimum, but a consequence of three aspects: (1) the particular
analytical structure of the
parametrization; (2) the characteristics of the dataset (in terms of the energy);
(3) the correlation among all the free fit parameters.
In Tables \ref{t1} and \ref{t2} we illustrate these correlations in the case of DF$_1$ and V1$_1$ and also
V4$_1$ and V6$_1$, respectively.

\vspace{0.3cm}

\subsection{Prediction at the 7 TeV Total $pp$ Cross Section}

\vspace{0.3cm}

In this section of BH we have identified some inconsistencies in the evaluation of the
quoted quantities, leading to misleading statements and conclusions.
As in the last subsection, we shall first present some introductory remarks and then treat each one of the 
three paragraphs.

\vspace{0.3cm}

$\bullet$ Introductory remarks

\vspace{0.3cm}

The TOTEM result at 7 TeV constitute a high-precision measurement, 
with quite small uncertainty if compared with any other high-energy experimental result
in the accelerator region \cite{totem}:

\vspace{0.3cm}

\centerline{$\sigma_{tot}^{pp}$ = 98.3 $\pm$ 0.2$^{\mathrm{stat}}$ $\pm$ 2.8$^{\mathrm{syst}}$ mb}

\vspace{0.3cm}

\noindent
or, adding the uncertainties in quadrature,

\vspace{0.3cm}

\centerline{$\sigma_{tot}^{pp}$ = 98.3 $\pm$ 2.8 mb}

\vspace{0.3cm}

Therefore, we understand that any acceptable description
of this datum demands agreement within the associated uncertainties (datum and prediction,
or fit result).
Unfortunately, as a first step in the research, we did not evaluate the
uncertainty regions in FMS (what has been done in FMS2). However,
in this reply, we shall provide additional information on this respect.
The error propagation
from the fit parameters have been based on the quoted references, namely \cite{root,bev}.
The standard error propagation includes variances and covariances associated
with all fit parameters \cite{bev} and is obtained from the error matrix
in the MINUIT code (available to any interested reader by request to Silva
at precchia@ifi.unicamp.br).

\vspace{0.3cm}

$\bullet$ First Paragraph - \textit{first two sentences}

\vspace{0.2cm}

- Commentary

Block and Halzen compare our curves in Fig. 5 stating that the result with V4
from Table 3 ($\sqrt{s}_{max}$ = 1.8 TeV ``goes slightly inside the lower error bar of the
plotted Totem result."

\vspace{0.2cm}

- Our reply

From our introductory remarks we do not consider this result in agreement with the
TOTEM datum. See also what follows.

\vspace{0.3cm}

$\bullet$ First Paragraph - \textit{last two sentences}

\vspace{0.2cm}

- Commentary

Block and Halzen state that, from the V4 parameters in Table 3 ($\sqrt{s}_{max}$ = 1.8 TeV) 
in a standard error evaluation, they find 
\begin{equation*}
\sigma_{tot} = \textbf{96.2} \pm 4.5 \text{ mb}.
\end{equation*}
and conclude that the result ``is in excellent agreement with the experimental value of
98.1 $\pm$ 2.3 found by Totem."

\vspace{0.2cm}

- Our reply

First, the TOTEM result is 98.3 $\pm$ 2.8 mb (as quoted above) and not 98.1 $\pm$ 2.3 mb
as referred to in BH.
Second, from the V4 parameters in Table 3 ($\sqrt{s}_{max}$ = 1.8 TeV) and including the variances
and covariances, the correct result is
\begin{equation*}
 \sigma_{tot} = \textbf{95.3} \pm 1.7 \text{ mb},.
\end{equation*}
which does not describe the TOTEM datum adequately. In particular, the above 
central value does not reach the lower extreme in the TOTEM uncertainty, namely 95.5 mb.
Without the covariances we obtain
$\sigma_{tot} = 95.3 \pm 5.4$ mb.

The quoted value given by Block and Halzen (96.2 $\pm$ 4.5 mb) \textit{does not correspond} to
the V4 parameters in Table 3 ($\sqrt{s}_{max}$ = 1.8 TeV), but to
the V4 parameters in
Table 4 ($\sqrt{s}_{max}$ = 7 TeV). Including the covariances, this value reads
96.2 $\pm$ 1.5 mb.
Therefore we are faced here with  serious errors in the evaluations
by Block and Halzen, leading to a misleading interpretation and conclusion.

\vspace{0.3cm}

$\bullet$ Second Paragraph

\vspace{0.2cm}

- Commentary

The reference here is on our DF result in Table 1 ($\sqrt{s}_{max}$ = 1.8 TeV and $\gamma$ = 2 fixed).
According to their evaluation, at 7 TeV,

\vspace{0.2cm}

\ \ \ \ \ \ \ \ \ \ \ \ \ \ \ \ \ \ \ \ \ \ \ \ \ \ \ \ \ \ \ \ \
$\sigma_{tot}$ = 95.4 $\pm$ 3.7 mb (1$\sigma$ diagonal error from $\beta$)

\vspace{0.2cm}

\ \ \ \ \ \ \ \ \ \ \ \ \ \ \ \ \ \ \ \ \ \ \ \ \ \ \ \ \ \ \ \ \
$\sigma_{tot}$ = 95.4 $\pm$ 8.8 mb (1$\sigma$ diagonal error in $s_h$ included)

\vspace{0.2cm}

\noindent
leading them to conclude: ``... DF model is also in good agreement with the Totem cross section, 
98.1 $\pm$ 2.3 mb."

\vspace{0.2cm}

- Our reply

Beyond the already mentioned error in the TOTEM result, taking into account the full error matrix
(variances and covariances) the correct prediction reads

\vspace{0.2cm}

\ \ \ \ \ \ \ \ \ \ \ \ \ \ \ \ \ \ \ \ \ \ \ \ \ \ \ \ \ \ \ \ \
$\sigma_{tot}$ = 95.4 $\pm$ 1.8 mb 

\vspace{0.2cm}

\noindent
The central value lies below the lower extreme in the TOTEM uncertainty, namely 95.5 mb and the
uncertainty region ($\pm$ 1.8 mb) barely reach the lower error bar.
We see no agreement of this result with the TOTEM
datum, 98.3 $\pm$ 2.8 mb.

\vspace{0.3cm}

$\bullet$ Third Paragraph

\vspace{0.2cm}

- Commentary

``As in the preceding Section, we find that the FMS $\gamma =$ 2 fits at low
energy, after allowing for errors in the predictions due to the statistical
errors in the fitting parameters, successfully predict the Totem total cross
section at 7 TeV, thus negating the necessity for considering a violation of
the Froissart bound. In simpler words, the FMS fits are consistent with a
saturated Froissart bound when the Totem point is included."

\vspace{0.2cm}

- Our reply

First, in our opinion, the uncertainties quoted by Block and Halzen
have been estimated in a ``rather cumbersome" way (compare
the procedures in the previous paragraphs 1 and 2). The standard error propagation \cite{bev}
include all variances and covariances given by the MINUIT error matrix.
Since covariances can assume negative values, the resulting
uncertainty may be smaller than the one obtained by taking into account only
the variances (diagonal errors). Second, from our arguments
on the above paragraphs and those in the last section, we can not agree
with the statement
``In simpler words, the FMS fits are consistent with a
saturated Froissart bound when the Totem point is included."

\section{Discussion}
\label{s3}

In this section we discuss some aspects of the FMS analysis and further information
that can be of interest. A short critical commentary on the BH analysis and results 
is also presented.

\subsection{On the FMS Analysis and Results}

The FMS analysis was motivated by the recent theoretical work by Ya. I. Azimov \cite{azimov1}
(see also \cite{azimov2,azimov3}), together with the well known results, obtained
a long time ago,
by Almaldi \textit{et al.} \cite{amaldi} and the UA4/2 Collaboration \cite{ua42}, which have indicated

\vspace{0.2cm}

\centerline{$\gamma$ = 2.10 $\pm$ 0.10 $\quad$ and $\quad$ $\gamma$ = 2.25$^{+0.35}_{-0.31}$}

\vspace{0.2cm}
\noindent
respectively. As commented in FMS, Azimov arguments that it is not obvious if Martin's derivation,
in the context of axiomatic local quantum field theory,
can be directly applied to hadronic processes (QCD). He discusses that ``under different assumptions about 
asymptotic behavior of nonphysical amplitude, the total cross section could grow even faster than
$\ln^2{s}$" \cite{azimov1}. Moreover, ``Increase of the total cross section faster than the 
log-squared energy does not mean violation of unitarity and is
\textit{not forbidden} by any general principles, contrary to a widespread opinion" \cite{azimov2}.

These have been the reasons why we have referred to the Froissart-Martin bound, namely the
log-squared bound. However, the  novel aspect of FMS consisted in \textit{only 
two phenomenological points}:

\begin{enumerate}

\item
the first inclusion of the 7 TeV TOTEM result for $\sigma_{tot}$ in the dataset to be reduced;

\item
the investigation of some consequences of this inclusion in what concern a $\ln^{\gamma}[s/s_h]$
leading high-energy contribution.

\end{enumerate}

Our analysis has indicated that, with the Amaldi \textit{et al}. parametrization
and the above inclusion, the TOTEM datum
cannot be adequately described in the case of $\gamma$ = 2 (which corresponds to the COMPETE
highest-rank parametrization). The 2012 analysis by the PDG, including the TOTEM
point, corroborates this indication \cite{pdg12}. 

Some explanation on this important point is in order.
The \textit{prediction} by the COMPETE Collaboration from the 2002 analysis, 
with their highest-rank parametrization ($\gamma$ = 2) \cite{compete1,compete2}, shows agreement with
the TOTEM datum at 7 TeV \cite{totem}.
However, that does not mean
the COMPETE highest-rank parametrization can describe this point \textit{once it is included in the
dataset}. That is exactly what the 2012 PDG result has indicated, corroborating
the conclusion previously presented in FMS.
Note that the 2012 PDG edition (Table 46.2 in \cite{pdg12}) quotes both
the Azimov article \cite{azimov1} and FMS (the reference, however, corresponds
to \cite{fm12}).

To illustrate the above point, we display in Figure 1 three results obtained through the
COMPETE highest-rank parametrization (Eq. (1) with $\gamma$ = 2).
The curves have been obtained with the values of the parameters from 

\vspace{0.2cm}

\noindent
- the 2002 COMPETE analysis (Table VIII in \cite{compete2}); 

\vspace{0.1cm}

\noindent
- the 2010 PDG edition (Table 41.2 in \cite{pdg10}), used as input in FMS;

\vspace{0.1cm}

\noindent
- the 2012 PDG edition \cite{pdg12} (Table 46.2 in \cite{pdg12}), which includes the
TOTEM datum in the dataset.

\vspace{0.2cm}

\begin{figure}[pb]
\centering
\epsfig{file=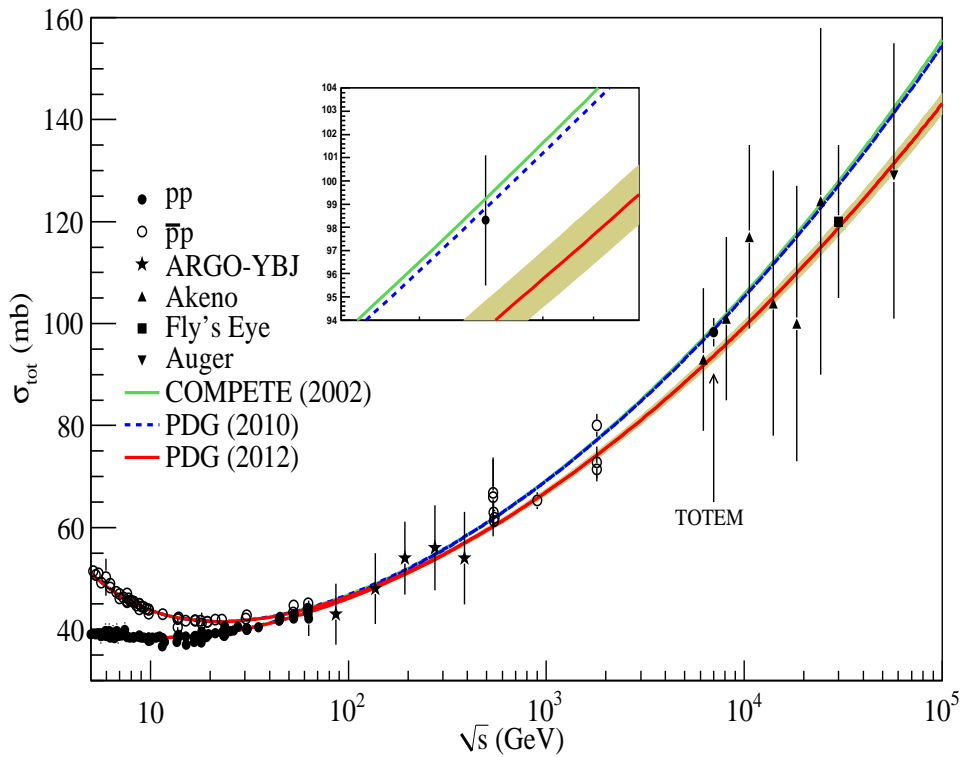,width=14cm,height=11cm}
\caption{Results for the $pp$ and $\bar{p}p$ total cross section with the COMPETE highest-rank parametrization
($\gamma$ = 2) from the 2002 COMPETE analysis \cite{compete1,compete2},
the 2010 PDG version \cite{pdg10} and the 2012 PDG version \cite{pdg12} (which includes the 7 TeV TOTEM datum
in the fitted dataset).}
\label{f1}
\end{figure}

\noindent
The values of the corresponding parameters, extracted from the above quoted references,
are displayed in Table \ref{t3}. The uncertainty region in Fig. \ref{f1}, in the case of the
2012 PDG result, has been determined through propagation of the errors in the parameters
(fourth column in Table \ref{t3}).
\begin{table}[ht]
 \centering
\caption{Fit results from the quoted references through parametrization (1)
with $\gamma$ = 2.
The parameters $a_1$, $a_2$, $\alpha$ and $\beta$
are in mb, $s_l$ and $s_h$ in GeV$^{2}$ and $b_1$, $b_2$ are dimensionless.}
\begin{tabular}{|c|c|c|c|}\hline
          & COMPETE 2002    & PDG 2010              & PDG 2012         \\
          & \cite{compete2} & \cite{pdg10}          & \cite{pdg12}     \\
\hline
$a_1$   \  &\ 42.1 \  $\pm$\ 1.3   \  &\ 42.53\  $\pm$\ 1.35  \  & 12.72 \ $\pm$\ 0.19  \  \\
$b_1$   \  &\ 0.467\  $\pm$\ 0.015 \  &\ 0.458\  $\pm$\ 0.017 \  & 0.462 \ $\pm$\ 0.002 \  \\
$a_2$   \  &\ 32.19\  $\pm$\ 0.94  \  &\  33.34\ $\pm$\ 1.04  \  & 7.35  \ $\pm$\  0.08 \  \\
$b_2$   \  &\ 0.5398\ $\pm$\ 0.0064\  &\  0.545\ $\pm$\ 0.007 \  & 0.550 \ $\pm$\ 0.005 \  \\
$s_l$   \  &\ 1 (fixed) \          \  &\  1 (fixed)\          \  &  16.21\ $\pm$\  0.16 \  \\
$\alpha$\  &\ 35.83\  $\pm$\ 0.40  \  &\ 35.35\  $\pm$\ 0.48  \  & 34.71 \ $\pm$\ 0.15  \  \\        
$\beta$ \  &\ 0.3152\ $\pm$\ 0.0095\  &\ 0.308\  $\pm$\ 0.010 \  & 0.265 \ $\pm$\ 0.050 \  \\
$s_h$   \  &\ 34.0\   $\pm$\ 0.54  \  &\ 28.9 \  $\pm$\ 5.4   \  & 16.21 \ $\pm$\ 0.16  \  \\
\hline
\end{tabular}
\label{t3}
\end{table}

From Figure 1, the \textit{predictions} by the COMPETE 2002 and PDG 2010
are in agreement with the TOTEM result, but not the data reduction
from the PDG 2012, which includes this point in the updated dataset: the curve and uncertainty region
lie below the TOTEM lower uncertainty bar, \textit{corroborating}, therefore,
\textit{the results and conclusions previously presented in FMS}.
Note also that these results (Figure 1) corroborate another conclusion in FMS, relating
the TOTEM point with the highest cosmic-ray estimates for the $pp$
total cross section (Fly's Eye Collaboration \cite{fly} and Pierre Auger Collaboration \cite{auger}).
In fact,
in what concerns our fit results with the Amaldi \textit{et al.} parametrization (and in the particular
case of Figure 1 with $\gamma$ = 2, namely the COMPETE highest-rank result), there is no agreement among these three points:
curves in consistency with the TOTEM datum lie above the central values of
the cosmic-ray estimations and the same is true in the inverse sense.

Based on all these facts, we understand that, once included in the dataset,
it is, at least, not obvious the 7 TeV TOTEM result can be described by a standard $\ln^{2}[s/s_h]$
leading dependence and that was the essential ``discovery" in FMS.
From our analysis, this effect is  related with the fundamental correlation
between $\gamma$ and the scale factor $s_h$ (and also with the subtraction constant $K$ in case
of simultaneous fits to $\sigma_{tot}$ and $\rho$ data: see Appendix A in FMS2 on the correlation
matrices). Note, from Table \ref{t3}, that in the COMPETE case, $s_h \sim$ 34 GeV$^2$ and in
the PDG 2012 edition, $s_h \sim$ 16 GeV$^2$ (below, therefore, $s_{min} =$ 25 GeV$^2$):
the effect of these differences can be seen in Fig. 1.
In particular, \textit{once included in the dataset}, it is not expected the 7 TeV TOTEM datum 
might be described for fixed $\gamma$ = 2 and $s_h = 2m_p^2$, as is the case in BH.

The TOTEM Collaboration has already obtained  three new 
high-precision measurements of the total cross section at 7 TeV, through different methods and techniques
\cite{totema}.
All the measurements are consistent within their uncertainties and
therefore confirm the first result they have obtained (which has been used in FMS and FMS2). 
In this respect, we can advance that the inclusion of these three  points in our dataset
leads also to solutions with $\gamma$ greater than two \cite{fms3,fms4},
corroborating the conclusions in FMS. However, the asymptotic ratio
between elastic/total-cross-section, discussed in FMS2, is still under investigation.
A luminosity-independent measurement at 8 TeV has been also reported,
indicating $\sigma_{tot} =$ 101.6 $\pm$ 2.9 mb \cite{totemb}. This value lies above the
prediction in FMS2, namely 98.7 $\pm$ 1.0 mb, for $\gamma$ = 2.346 $\pm$ 0.013
and $s_h$ = 0.383 $\pm$ 0.041 GeV$^2$ \cite{fms2}.

At this point, we could conjecture (if not speculating) on the implication
of a possible increase of $\sigma_{tot}$ faster than $\ln^2{s}$.
In contrast with an effective violation of the Froissart-Martin bound,
a fast rise of the total cross section might also be associated
with some local effect at the LHC energy region, so that, asymptotically, the  bound remains valid.
A faster-than-squared-logarithm rise points also to the possibility of a power-like
behavior \cite{dl1,dl2,dl3}, which has always been and important and representative approach
(see, for example, the unitarized model \cite{kope1,kope2}).
These conjectures are not in disagreement with the recent theoretical
arguments by Azimov  \cite{azimov1,azimov2,azimov3}.

\subsection{Brief Commentary on the BH Analysis and Results}

Here we present some critical comments on the BH parametrization, the Aspen Model
and the BH results for the total, elastic and inelastic cross sections.

\subsubsection{Analytical Parametrization}

The BH analytical parametrization for the total cross sections, used in 
global fits to $\sigma_{tot}$ and $\rho$-values from $pp$ (+)
and $\bar{p}p$ (-) scattering, reads \cite{bhcomm}

\begin{eqnarray}
\sigma^{\pm}(\nu) =  
\beta_{P'}\, \left[\frac{\nu}{m}\right]^{\mu - 1}  
\, \pm \, \delta \, \left[\frac{\nu}{m}\right]^{\alpha - 1} +
c_0\, + c_1\, \ln\left(\frac{\nu}{m}\right)\, +
c_2\, \ln^2 \left(\frac{\nu}{m}\right),
\end{eqnarray}
\label{eq2}
where $\nu$ is the laboratory energy, $m$ is the proton mass and, in terms
of the c.m. energy, $\nu/m \approx s/2m^2$.

Comparison with the parametrization used in FMS, Eq. (1), shows that, for 
$\gamma$ = 2 (fixed), both forms have the same \textit{analytical structure}.
However, the striking difference between FMS and BH approaches concerns the
number of fixed and free parameters and mainly, in the BH case, the way the
parameters are fixed, as discussed in what follows.

In the FMS analysis, except for a dimensional notation choice, namely $s_l$ = 1 GeV$^2$, 
all the 8 parameters involved have been treated as free or fixed in the 6 variants
considered. That is, there is no \textit{ad hoc} fixed parameters, except for 
particular variant tests. In all cases, the uncertainties in the free
parameters have been explicitly given. Moreover, by letting free
different parameters in a \textit{nonlinear}
fit, we are able to investigate all the correlations involved, the
variances, covariances and, as a consequence, the global uncertainties
in all fitted and predicted quantities, as done in FMS2.

In the BH approach, besides the arbitrary fixed energy scale (corresponding
to $s_h = 2m^2$ at both low and high energy regions), among the 7 parameters
in Eq. (2),
5 are fixed (without uncertainties) and only 2 are free in
data reductions. Specifically, from Table III in \cite{bh05}:

\vspace{0.1cm}

- Fixed parameters: 
$\beta_{P'}$ = 31.10 mb, $\mu$ = 0.5, $\delta$ = - 28.56 mb,
$\alpha$ = 0.415 and $c_0$ = 37.32 mb;

\vspace{0.1cm}
- Fitted parameters: $c_1$ = - 1.440 $\pm$ 0.070 mb and $c_2$ = 0.2817 $\pm$ 0.0064 mb.

\vspace{0.1cm}

As a consequence, the parametrization is linear in any reduction to $\sigma_{tot}$ data, leading to 
unique solution \cite{bev}. That, obviously, contrasts with Eq. (1) and with the strategies
in FMS and FMS2, as commented above.

In BH  the Reggeons intercepts are fixed, corresponding to
$b_1$ = 0.5 and $b_2$ = 0.585 in Eq. (1), which is not in agreement with the spectroscopic
data (Chew-Frautschi plots) and scattering fit results, as obtained by several authors
\cite{compete1,compete2,pdg10,pdg12,cmg,ckk,lm,lmm,lmmbjp}. Since this assumption permeated the 
intermediate and low  energy region,
in our opinion, it puts some limits on the reliability of a formal connection with
the Finite Energy Sum Rules at low energy.
The fixed parameters do not allow the study of correlations and
their effects in the fitted and, most importantly, in the predicted quantities.
We shall return to this point in the next subsection on the Aspen model.

We also note that, although (1) and (2) have the same analytical structure,
the BH and FMS high energy formulations for $\gamma$ = 2 are not equivalent,
even in the cases (variants) with $b_1$ and $b_2$ fixed. In fact, in FMS all
the other parameters are free and the fit extends \textit{simultaneously}
to both low- and high-energy data ($\sigma_{LE}$ and $\sigma_{HE}$ contributions
in FMS). As a consequence there is strong correlations among the parameters
from both $\sigma_{LE}$ and $\sigma_{HE}$ (see, for example, the V4$_1$ coefficients
in Table II in the case of $a_1$ and $s_h$, $\alpha$ and $\beta$). Since
that is not the case in BH formulation ($\beta_{P'}$ and $\delta$,
corresponding to $a_1$ and $a_2$, are fixed), we see no correspondence between
the high energy formulations, as stated in BH (after Eq. (7) in that paper).

In the BH approach the main hypothesis concerns the imposing that ``the fits 
to the high energy data
smoothly join the cross section and energy dependence obtained by averaging 
the resonances at low energy" \cite{bh05}. In the FMS analysis, on the contrary,
``We have tried to identify possible high-energy effects that may be unrelated to
the trends of the lower-energy data..." \cite{fms1}.
Therefore, the assumptions, approaches and strategies in FMS and BH are completely
different. We see no reason for the comparative discussion presented in BH.

\subsubsection{Aspen Model}

In the Aspen model \cite{asp1,asp2} two fundamental quantities, the mass scale $m_0$
and the coupling constant $\alpha_s$, are unknown parameters, fixed to
\textit{ad hoc} values of 600 MeV and 0.5, respectively in order to obtain best fits
in data analysis. The $\epsilon$ parameter, from the gluon structure function,
is also fixed at 0.05.

The two fundamental parameters, $m_0$ and $\alpha_s$, have been reinterpreted by Luna \textit{et al.}
in the context of a Dynamical Gluon Mass (DGM) approach \cite{lmmmn1,lmmmn2}.
That has allowed  connections with nonperturbative QCD,
as expected in the soft sector represented by the elastic scattering processes.
In the DGM approach, two essential parameters are the dynamical gluon mass scale $m_g$
and the soft Pomeron intercept $\epsilon$. More recently and most important for our purposes,
Fagundes, Luna, Menon and Natale (FLMN) have developed a detailed analysis on the influence
in the evaluated quantities associated with physical intervals
for the $m_g$ and $\epsilon$ parameters \cite{flmn1,flmn2}. Moreover,
in a similar way as done recently by Achilli \textit{et al.} \cite{achilli}, FLMN have established
bounds and uncertainty regions in all evaluated quantities (fitted and predicted), in accordance with the 
relevant physical intervals for $m_g$ and $\epsilon$. The main conclusion is that the
uncertainty regions play a crucial role in the energy region above that
used in the data reductions. In other words, the relevant intervals for each parameters affect substantially
the high-energy predictions and they cannot be fixed at \textit{ad hoc} values, without a 
clear physical justification.

The above conclusion, expressed in \cite{flmn2}, puts serious limits
on the predictions of phenomenological models constructed on the basis of fixed parameters,
whose numerical values do not have an explicit justification and/or whose \textit{consequences in the evaluated quantities
are not investigated or even discussed}.
In our opinion, these limitations are present in the foundations of both the BH analytical
parametrization and the Aspen model.

\subsubsection{Experimental Evidence}

In order to connect their global analytical (empirical) fits to $\sigma_{tot}$ and $\rho$ data with
the inelastic cross section, $\sigma_{inel}$, Block and Halzen use the 
predictions from the Aspen model,
in a kind of hybrid approach (semi-empirical or perhaps semi-phenomenological) \cite{bhcomm,bh12}. 
The model prediction is parametrized by an analytic expression (Eq. (8) in \cite{bhcomm}), with fixed
mass and power parameters, without any reference to the uncertainties
in the fit parameters. With this hybrid approach, from the model evaluation of $\sigma_{inel}$ and the
fit result for $\sigma_{tot}$ they infer that
\begin{eqnarray}
\frac{\sigma_{inel}(s)}{\sigma_{tot}(s)}\  \rightarrow \ \mathrm{0.509} \pm \mathrm{0.021}
\qquad \mathrm{as} \qquad
s \rightarrow \infty.  
\nonumber
\end{eqnarray}
a result statistically consistent with the black-disc limit.

As commented in FMS2, we understand that the least ambiguous way to estimate the
inelastic cross section is through the s-channel unitarity,
$\sigma_{inel}$ = $\sigma_{tot}$ - $\sigma_{el}$, as has been done by the TOTEM Collaboration
\cite{totem}. That avoids the model dependence involved in direct estimation of
$\sigma_{inel}$, due to single and double diffraction contributions.

In this respect, it may be interesting to note the prediction in BH for the
inelastic cross secion at 7 TeV (Fig. 3 in \cite{bhcomm}): the uncertainty region barely reaches the 
lower error
bar in the TOTEM result. Moreover, for the total cross section, the central value of the prediction
reaches the low extreme bar of the TOTEM result (and the uncertainty region, above the central value,
lies through the lower error bar of the TOTEM point). It should be also noted that as a one-channel eikonal
formalism, the Aspen model does not take explicit account of the inelastic diffractive contributions.

In what concerns the BH total cross section prediction  at 57 TeV and the recent
estimation of this quantity by the Pierre Auger Collaboration \cite{auger},
we stress a peculiar statement in \cite{bhcomm} (also present in \cite{bh12}):

\begin{quote}
``In particular, the agreement with the new highest energy (57 TeV) experimental measurement of both
$\sigma_{tot}$ and $\sigma_{inel}$ is striking."
\end{quote}

First, certainly these are not \textit{experimental measurements}, due to the strong model dependence involved,
as already discussed by several authors and also in \cite{auger}. They constitute estimations of these
quantities because they are based on extrapolations from models tested only in lower energies.
Moreover, recall that the estimation for the $pp$ total cross section
at 57 TeV reads \cite{auger}

\vspace{0.2cm}

\centerline{
$
\sigma_{tot}^{pp} = [ 133 \pm 13\, \mathrm{(stat)}^{+ 17}_{-20}\,\mathrm{(sys)} 
\pm 16\, \mathrm{(Glauber)}]\, \mathrm{mb}
$.} 

\vspace{0.2cm}

\noindent
As discussed in FMS2 (and also in our section II.B), systematic and, here, theoretical (Glauber)
uncertainties are equally probable quantities (do not follow a Gaussian distribution). That means
the above central value is equally likely to lie in any place limited by the corresponding
nonstatistical uncertainties, namely around $\pm$ \ 23 mb. Therefore, we see no physical meaning
in a statement referring to a  ``striking agreement".

At last, we understand that the BH analysis and the Aspen model represent important contributions
in the investigation of the high-energy elastic hadron
scattering. However, given the model character and \textit{ad hoc} assumptions involved
we see no conclusive evidence that the BH results constitute an unique 
and exclusive solution for the rise of the total cross section at high energies.
In this respect, a reanalysis by Block and Halzen, including in their dataset
the four high-precision TOTEM measurements and presenting their linear fit prediction
at higher energies, might be instructive.

\section{Conclusions and Final Remarks}
\label{s4}

In FMS we have presented a study on the rise of the total hadronic cross section, 
with focus on the recent 7 TeV TOTEM result. The analysis was based on a specific 
class of analytical parameterization, Eq. (1). Since the effects of all
parameters involved have been considered, we were faced with a nonlinear data 
reduction, which constitutes a non-trivial problem.

Our strategy in FMS (and FMS2) 
has been to investigate a grid of different
physical choices for feedbacks and the corresponding solutions. In both FMS and FMS2, beyond a second distinct
possibility,
we have considered ``conservative" choices for the initial values, namely results previously
obtained by the PDG with $\gamma$ = 2.
The data reductions have been developed through standard
running of the MINUIT, namely the default MINUIT error analysis.
Perhaps, in the fits presented in FMS a true minimum had not been reached in some cases.
However, we understand that this does not invalidate
 the \textit{general and global conclusions} from our ``grid strategy" approach (summarized 
in section II and, the corresponding results, in
Fig. 8 of FMS).
Anyway, further analysis, looking for optimizations in the use of the MINUIT code,
including also tests with other computational tools for nonlinear fits (as the 
subroutine MINOS \cite{minuit2}), are certainly important and we intend to implement that.

The lack of unique solutions in our nonlinear fit procedures has been 
referred to in FMS and FMS2. We have never claimed  to have obtained \textit{unique or absolute solutions}.
In particular, we have concluded in FMS (the equations refer to that paper),

\begin{quote}
``From our data reductions through parametrization (3-5) to $pp$ and $\bar{p}p$ scattering above
5 GeV, including the 7 TeV TOTEM result, we conclude that the total hadronic cross 
section may rise faster than $\ln^2{s}$ at high energies."
\end{quote}

\begin{quote}
``Our results suggest that the energy dependence of the hadronic total cross section at high energies still
constitutes an open problem."
\end{quote}

\noindent
and in FMS2 we have stressed,

\begin{quote}
``We also emphasize that
our results represent possible consistent statistical solutions for the behavior of the
total cross section, but do not correspond to unique solutions."
\end{quote}

It is important to note that at the highest energies, once treated as free fit parameters,
the exponent $\gamma$ and the scale factor $s_h$
are correlated in nonlinear data reductions.
In the lack of formal and/or theoretical information on the value of this
scale factor and without the \textit{canonical assumption} $\gamma$ = 2, different solutions,
in agreement with the experimental data, can 
be obtained for the leading $\ln^{\gamma}(s/s_h)$ contribution.
Therefore,
we understand that the analytical description of the rise of the total hadronic cross section
at high energies still constitutes an open problem, demanding  further investigations. 
New high-precision data to be published and to be obtained by the TOTEM Collaboration at 8 -
 14 TeV  are expected to shed light on the subject.

At last,
based on both the ``$commentary$" by Block and Halzen and this ``$reply$",
we see no evidence that invalidate the analysis, results, conclusions and the honest statements 
expressed in FMS.

\section*{Acknowledgments}

We are thankful to J.R.T. de Mello Neto for useful
discussions.
Research supported by FAPESP 
(Contracts Nos. 11/15016-4, 11/00505-0, 09/50180-0).

\end{document}